\documentclass[aps,prl,twocolumn,preprintnumbers,showpacs,superscriptaddress,nofootinbib,floatfix,10pt]{revtex4-1}
\pdfoutput=1
\usepackage{textcomp}
\usepackage[english]{babel}
\usepackage{amsmath,amssymb,amsbsy,booktabs}
\usepackage{bm}
\usepackage{color}
\usepackage{array}
\usepackage{amstext}
\usepackage{graphicx}
\usepackage{amsfonts}
\usepackage{bm}
\usepackage{dcolumn}
\usepackage{rotating}
\usepackage{epstopdf}
\usepackage{esint}
\usepackage{url}
\usepackage[colorlinks=true,
linkcolor=blue,
filecolor=blue,
anchorcolor=blue,
urlcolor=blue,
citecolor=blue
]{hyperref}

\usepackage[utf8]{inputenc}

\allowdisplaybreaks[0]

\begin{document}
	
	\title {\bf Probing New Physics Signals with Symmetry-Restored Yukawa Textures}
	\author{Shao-Ping Li}
	\email{ShowpingLee@mails.ccnu.edu.cn}
	\affiliation{Institute of Particle Physics and Key Laboratory of Quark and Lepton Physics~(MOE),\\
		Central China Normal University, Wuhan, Hubei 430079, P.~R.~China}
	
	\author{Xin-Qiang Li}
	\email{xqli@mail.ccnu.edu.cn}
	\affiliation{Institute of Particle Physics and Key Laboratory of Quark and Lepton Physics~(MOE),\\
		Central China Normal University, Wuhan, Hubei 430079, P.~R.~China}

\begin{abstract}
We implement a symmetry violation guideline into a two-Higgs-doublet model embedded with three right-handed neutrinos, and exploit the generic Yukawa structures of the model via a hypothetical symmetry restoration of a global $U_Q(1)^3\times U_L(1)^3$ symmetry. We then apply a mass-powered parametrization to construct the phenomenologically motivated Yukawa interactions, which enables us to incorporate correlatively the neutrino mass, dark matter, as well as the lepton-flavor universality violations in $R_{K^{(*)}}$ and $R_{D^{(*)}}$.  Specifically, two atmospheric-scale neutrino masses are generated by a low-scale seesaw mechanism, while the much lighter one, being of $\mathcal{O}(10^{-6})~{\rm eV}$, is fixed by a $7.1~{\rm keV}$ sterile neutrino dark matter produced primordially by the freeze-in mechanism. On the other hand, the neutrino and the charged-lepton mass hierarchies encoded in the mass-powered textures can naturally account for the $R_{K^{(*)}}$ and the $R_{D^{(*)}}$ anomalies, respectively. As a further application, a milder discrepancy of the muon $g-2$ is observed, which has also been implied by the recently refined lattice results.
\end{abstract}


\maketitle

\textit{Introduction}.---The neutrino mass and dark matter (DM) are two well-known signatures that require new physics (NP) beyond the Standard Model (SM). Towards solving these two momentous issues in particle physics, some tantalizing NP signals observed in B-meson decays, such as the ratios of the branching fractions 
\begin{align}
R_{K^{(*)}}=\frac{\mathcal{B}(B \rightarrow K^{(\ast)} \mu^+ \mu^-)}{\mathcal{B}(B\rightarrow K^{(\ast)} e^+ e^-)}
\end{align}
and 
\begin{align}
R_{D^{(*)}}=\frac{\mathcal{B}(B\rightarrow D^{(\ast)} \tau \bar{\nu})}{\mathcal{B}(B\rightarrow D^{(\ast)} e(\mu)\bar{\nu})},
\end{align}
 can provide complementary incentives in pinning down the underlying NP theory~\cite{Bifani:2018zmi,Ciezarek:2017yzh}.

With the updated LHCb measurement in the dilepton invariant mass squared range $q^2=[1.1,6.0]~{\rm GeV}^2$~\cite{Aaij:2019wad},
\begin{align}
R_{K}=0.846^{+0.060}_{-0.054}(\text{stat})^{+0.016}_{-0.014}(\text{syst}),
\end{align} 
together with the previous $R_{K^{*}}$ data~\cite{Aaij:2017vbb}, 
\begin{align}
R_{K^\ast}=0.69^{+0.11}_{-0.07}{\rm (stat)} \pm 0.05{\rm (syst)},
\end{align} 
discrepancies at the level of $\sim2.5\sigma$ are found with respect to the corresponding SM predictions (both being equal to one up to a few percent corrections~\cite{Bordone:2016gaq}). The new Belle result of $R_{K^*}$~\cite{Abdesselam:2019wac}, given its sizable uncertainty, is also compatible with the LHCb measurement~\cite{Aaij:2017vbb}. For the ratios $R_{D^{(*)}}$, the latest combination of the BaBar~\cite{Lees:2012xj,Lees:2013uzd}, Belle~\cite{Huschle:2015rga,Hirose:2016wfn,Hirose:2017dxl,Abdesselam:2019dgh}, and LHCb~\cite{Aaij:2015yra,Aaij:2017uff,Aaij:2017deq} measurements performed by the Heavy Flavor Averaging Group gives~\cite{HFLAV:2019}
\begin{align}
R_D=0.340\pm 0.027(\text{stat})\pm 0.013(\text{syst}),
\\
R_{D^*}= 0.295\pm 0.011(\text{stat})\pm 0.008(\text{syst}),
\end{align}
which, after taking into account their correlation of $-0.38$, exhibit a $3.1\sigma$ deviation from the SM predictions, 
\begin{align}
R^{\rm SM}_D=0.299\pm0.003,\quad
R^{\rm SM}_{D^\ast}= 0.258\pm 0.005.
\end{align}
See Ref.~\cite{HFLAV:2019} for a collection of refined SM predictions. Benefiting from cancellations of large parts of the hadronic uncertainties, the $R_{K^{(*)}}$ and $R_{D^{(*)}}$ anomalies, if confirmed, would hint at lepton-flavor universality violation (LFUV) that is not accountable within the SM.

In recent years, a two-Higgs-doublet model (2HDM) with generic Yukawa structures~\cite{Crivellin:2013wna} has been investigated intensively. This framework can explain the $R_{D^{(\ast)}}$ anomalies with sizable top-charm couplings~\cite{Crivellin:2015hha,Iguro:2017ysu,Li:2018rax} and, when embedded with three right-handed neutrinos, address the neutrino mass and the $R_{K^{(\ast)}}$ anomalies simultaneously~\cite{Li:2018rax}, or the $7.1~{\rm keV}$ sterile neutrino dark matter~\cite{Adulpravitchai:2015mna,Baumholzer:2018sfb}. One might, therefore, expect the general 2HDM with three right-handed neutrinos as a promising candidate to unify these NP signals. However, the model invokes generically overabundant (unknown) Yukawa parameters and hence limits its capabilities of theory prediction and correlation. Thus, understanding the origin of the phenomenologically motivated Yukawa structures becomes especially crucial to further exploit the model.

In this paper, we shall be concerned with the possible correlations among these generic Yukawa couplings in a 2HDM with three right-handed neutrinos (2HDM$+3N_R$). To this end, we consider here a hypothetical symmetry restoration of a global $U_Q(1)^3\times U_L(1)^3$ symmetry that should have been broken by the Yukawa couplings in the mass-eigenstate basis. By attributing the symmetry breaking sources to the known fermion masses (or equivalently the dimensionless Yukawa eigenvalues), we then find that the unknown Yukawa matrices of the model can be built out of these known fermion masses. To visualize this idea phenomenologically, we propose a simple mass-powered parametrization of these non-diagonal Yukawa matrices, which enables us to incorporate correlatively the neutrino mass, DM, as well as the $R_{K^{(\ast)}}$ and $R_{D^{(\ast)}}$ anomalies within a single framework.

\textit{The model}.---In many model buildings concerned with the Yukawa structures, the usual strategy is to implement symmetry invariance into the Yukawa sector by invoking some heavy dynamical fields, as adopted \textit{e.g.} in the Froggatt-Nielsen mechanism~\cite{Froggatt:1978nt} and the minimal flavor violation hypothesis~\cite{Buras:2000dm,DAmbrosio:2002vsn}. Such considerations are strongly supported by the low-energy flavor observations. Nevertheless, in the low-energy regime, there seems to be no exact but rather broken symmetries, or approximate symmetries with some random perturbations. Assuming that the broken symmetries are not recovered by any dynamical fields, we exploit here an interesting scenario that does not receive much attention. Explicitly, we shall assume that the Yukawa interactions have indeed completely and explicitly broken some family symmetries in the mass-eigenstate basis, but with limited and correlative perturbations, or symmetry breaking sources (SBS). Furthermore, the SBS are located only in the Yukawa sector such that, when the SBS vanish, the corresponding symmetries can be restored universally and completely throughout the Yukawa Lagrangian.
	
It should be emphasized that, if there are various unrelated SBS, we must tune these SBS synchronously to vanish, in order to make the symmetry restoration universal throughout the whole Lagrangian. This is technically feasible but not quite natural. In this context, we shall assume further that the various SBS have some common origins, so that the naturally synchronous vanishing of the SBS is triggered by these common origins. This prescription allows the Yukawa structures to be generated only by the common SBS, and hence reduces dramatically the number of free parameters. Obviously, such a setup for constructing the Yukawa textures is not based on the \textit{symmetry invariance} principle, but rather on the \textit{symmetry violation} guideline, which is concerned with how the flavor symmetries are broken completely by the Yukawa interactions in the mass-eigenstate basis, and can be introduced via the following criteria: 
\begin{itemize}
 \item[$(i)$.] The Yukawa interactions in the mass-eigenstate basis have broken explicitly some family symmetries but only via limited and correlative SBS.
		
 \item[$(ii)$.] The Yukawa textures generated by these SBS are subject to a hypothetical symmetry restoration. This indicates that the  corresponding symmetry will be recovered universally and completely throughout the whole Lagrangian under vanishing SBS.
		
 \item[$(iii)$.] There are no other adjustable free parameters beyond those SBS, so that the symmetry restoration is triggered only by vanishing SBS.
\end{itemize}
The universal symmetry restoration in criteria $(ii)$ corresponds to the natural setup in which all the symmetry perturbations can vanish synchronously when the common SBS origin is tuned to zero, while the criteria $(iii)$ forces the number of free parameters to be further reduced during the construction of the Yukawa matrices.
	
In the rest of our work, we shall apply these criteria in the 2HDM$+3N_R$ framework to construct the Yukawa interactions, which are found to be able to account for the NP signals considered. In the mass-eigenstate basis, the Yukawa interaction Lagrangian can be parametrized in a compact form as
\begin{align}\label{lag}
\mathcal L_{Y}&=\mathcal{L}_{H_1}+\mathcal{L}_{H_2},
\nonumber \\[0.5mm]
\mathcal{L}_{H_1}&=-\bar{Q}_L V^\dagger \hat{Y}_u \tilde{H}_1 u_R- \bar{Q}_L \hat{Y}_d H_1d_R- \bar{E}_L\hat{Y}_\ell H_1  e_R
\nonumber \\[0.2mm]
&\quad - \bar{E}_L \tilde{Y}_{\nu}\tilde{H}_1N_R + \rm H.c.,
\nonumber \\[0.5mm]
\mathcal{L}_{H_2}&=-\bar{Q}_L V^\dagger \mathcal{Y}_u \tilde{H}_2 u_R - \bar{Q}_L \mathcal{Y}_d H_2 d_R-\bar{E}_L\mathcal{Y}_\ell H_2  e_R
\nonumber \\[0.2mm]
&\quad - \bar{E}_L \mathcal{Y}_{\nu} \tilde{H}_2 N_R+ \rm H.c..
\end{align}
Here, $Q_L\equiv (V^\dagger u_L, d_L)^T$ and $E_L\equiv (U_\nu^* \nu_L, e_L)^T$, with $V$ and $U_\nu$ corresponding respectively to the quark and neutrino mixing matrices observed in experiments, no longer form the $SU_L(2)$ doublets. More importantly, we have particularly neglected, in Eq.~\eqref{lag}, the terms that are either proportional to the active neutrino mass or further suppressed by the light-heavy neutrino mixing parameter $R^*\simeq M_D M_R^{-1}=v \tilde{Y}_\nu M_R^{-1}/\sqrt{2} $~\cite{Ibarra:2010xw}. Note that the matrices $\hat{Y}_{u,d,\ell}=\sqrt{2} m_{u,d, \ell}/v$ are already in a diagonal form, but $\tilde{Y}_\nu$ is not. In addition, the matrices $\mathcal{Y}$ remain non-diagonal and encode the unknown Yukawa interactions arising in the 2HDM$+3N_R$ framework.

The two scalar doublets $H_{1,2}$ are given in the Higgs basis as~\cite{Branco:2011iw}
\begin{eqnarray}
H_1 =\left(
\begin{array}{c}
G^+\\
\frac{v+\phi_1+iG}{\sqrt{2}}
\end{array}
\right),~~~
H_2=\left(
\begin{array}{c}
H^+\\
\frac{\phi_2+iA}{\sqrt{2}}
\end{array}
\right),
\label{HiggsBasis}
\end{eqnarray}
with the vacuum expectation value $v\simeq 246~{\rm GeV}$. The neutral scalars $\phi_{1,2}$ are the superposition of the two mass eigenstates $H$ and $h$ via $\phi_{1(2)}=\cos\alpha(-\sin\alpha)H+\sin\alpha(\cos\alpha)h$, with the mixing angle determined by
\begin{align}\label{mixing angle}
\tan(2\alpha)=\frac{2\lambda_6 v^2}{\sqrt{(M_H^2-M_h^2)^2-4\lambda_6^2 v^4}},
\end{align}
where $\lambda_6$ is the quartic mixing coupling of the term $[(H_1^\dagger H_1)(H_1^\dagger H_2)+\rm H.c.]$ in the Higgs-basis scalar potential~\cite{Branco:2011iw}. Let us consider both $\mathbb{Z}_2$ and $CP$ symmetries that are conserved by the Higgs-basis scalar potential but violated by the Yukawa sector. In this case, $\lambda_6=0$ and hence the two neutral Higgs bosons ($H$ and $h$)   decouple from each other. Furthermore, choosing one of the solutions, $\alpha=\pi/2$, $h$ would return to the SM Higgs boson (the so-called alignment limit). As a consequence, there exist no tree-level flavor-changing neutral currents (FCNC) involving $h$, as is observed in the SM. 

It can be readily seen that, in the mass-eigenstate basis, Eq.~\eqref{lag} breaks  
a global $U_Q(1)^3\times U_L(1)^3$ symmetry, with the following transformation rules:
\begin{align}\label{symmetry transformation}
(u_L,d_L)_i &\rightarrow e^{i \mathcal{Q}_{i} \alpha_i} (u_L,d_L)_i,\;\;
u(d)_{Ri} \rightarrow e^{i \mathcal{U}(\mathcal{D})_i \alpha_i} u(d)_{Ri},
\nonumber \\[0.2cm]
(\nu_L,\ell_L)_i & \rightarrow e^{i \mathcal{L}_{i} \beta_i} (\nu_L,\ell_L)_i,\;\;
\ell(N)_{Ri} \rightarrow e^{i \mathcal{E}(\mathcal{N})_{i}  \beta_i} \ell(N)_{Ri},
\end{align}
where $\mathcal{Q}_i$, $\mathcal{U}_i$, $\mathcal{D}_i$, $\mathcal{L}_i$, $\mathcal{E}_i$ and $\mathcal{N}_i$ denote the corresponding $U(1)$ charges of the $i$-th fermion generation. The specific charge assignments are irrelevant, provided that no particular relations among these charges would change the broken symmetry. This makes it possible for us to consider the most general case where the transformation rules specified by Eq.~\eqref{symmetry transformation} exhibit a complete breaking of the $U_Q(1)^3\times U_L(1)^3$ symmetry.

In the following, we shall seek for the common SBS that satisfy the criteria given above. Let us take the neutral currents for a start. Considering the quark neutral currents (as well as the quark mass terms) of $\mathcal{L}_{H_1}$, one can see that the $U_Q(1)^i$ symmetry would be restored when the $i$-th quark masses hypothetically vanish, $m_{i}^{u,d}=0$, since these terms are all determined by the Yukawa matrices $\hat{Y}_{u,d}$. Based on this observation, we can conjecture the quark masses, or equivalently the dimensionless Yukawa eigenvalues $y^f_i$ defined by
\begin{align}\label{Yukawa seeds}
y^{f}_i y^{f}_j \delta_{ij} v=m^f_i,
\end{align}
as the common SBS. Turning now to the quark neutral currents of $\mathcal{L}_{H_2}$, the non-diagonal Yukawa matrices $\mathcal{Y}_{u,d}$ are also expected to be generated by the common SBS. Following the guideline specified by the three criteria given above, we can construct $\mathcal{Y}_{u,d}$ in the following form:
\begin{align}\label{mp}
 \mathcal{Y}^{f}_{ij}= (y^f_i)^{\tilde{n}^f_i}\times (y^f_j)^{\tilde{n}^f_{j}},
\end{align}
where $f=u,d$, and the factors $\tilde{n}^f_{i}$ denote the powers to which the generic Yukawa matrices $\mathcal{Y}^{f}$ are built out of the eigenvalues $y^f_i$. The reason for such a power realization is that these power factors, while being free, do not play the role of SBS. In fact, these powers can be readily understood by noting that the Yukawa matrices are built out of the eigenvalue products themselves, as required by the criterion $(iii)$. It is then readily seen that, when the $i$-th quark masses hypothetically vanish, or $y^{u,d}_i=0$, the corresponding $U_Q(1)^i$ symmetry would be restored throughout the neutral currents of Eq.~\eqref{lag}.

Next, let us consider the quark charged currents. In this part, due to the presence of quark mixing, the SBS role is now played by both the mixing matrix and the quark masses. To still retain the masses as the common SBS, some correlation should be established between the quark masses and the mixing matrix, such that the mixing would synchronously vanish when quarks are hypothetically massless. Since the general mass matrix $M_f$ is constructed in terms of the mass eigenvalues through a bi-unitary rotation, $M_f=V_L^{f\dagger} \hat{M}_f V_R^{f}$, and the combination of up- and down-quark rotations gives the physical CKM mixing matrix, $V=V_L^u\,V_L^{d\dagger}$, the same guideline allows us to build the general mass matrix with the following texture:
\begin{align}\label{mp2}
 M_f=v\, (y^f_i\times y^f_j)^{n_{ij}}.
\end{align}
Again, the factor $n_{ij}$ denotes the power to which the general mass matrix $M_f$ is expanded in terms of the Yukawa eigenvalues. Note that the general mass matrix cannot exhibit a form like Eq.~(13), otherwise the mass matrix would be of rank one and unacceptably render only one fermion generation massive, while this is not the case for $\mathcal{Y}_f$ as it is not responsible for fermion masses in the considered mass-eigenstate basis.

Eq.~\eqref{mp2} indicates that, if the general mass matrices in a non-physical basis were also constructed in terms of the Yukawa eigenvalues, the flavor spectra would be induced by the random powers (non-SBS parameters) $n_{ij}$ without affecting the broken symmetry pattern after changing to the physical basis. In addition, the synchronous vanishing of the physical mixing matrix triggered by fermion masses can now be seen as follows:
\begin{equation}\label{synchronization}
m^f_{i_0}(y^f_{i_0})=0\,\Rightarrow \,
\begin{cases}
(V^{f}_{L})_{i_0 j}=(V^{f}_{L})_{ji_0}=0,\, j\neq i_0,
\\[0.2cm]
(V^{f}_{L})_{i_0 i_0}=1,
\end{cases}.
\end{equation}
In this way, the hypothetically vanishing of quark masses as the common SBS is able to prompt a universal $U_Q(1)^{3}$ symmetry restoration throughout the Yukawa Lagrangian specified by Eq.~\eqref{lag}. In fact, the same restoration is also valid in the $W^\pm$-mediated charged currents due to Eq.~\eqref{synchronization}. 


The above analysis can be directly applied to the charged-lepton sector. However, it is non-trivial to see a hypothetical symmetry restoration in the neutrino sector. This can be seen in several ways. Firstly, unlike the Dirac fermions, the left-handed neutrinos do not combine with the corresponding right-handed counterparts to form the physical masses. Secondly, as the generation of active neutrino masses relies on the seesaw mechanism which entails a non-singular right-handed neutrino mass matrix $M_R$, turning hypothetically the right-handed Majorana neutrinos massless would cause conceptual issues for the active neutrino masses. Finally, as the seesaw mass formula, 
\begin{align}\label{seesaw}
 M_\nu\simeq -M_DM_R^{-1}M_D^T,
\end{align}
is a leading-order result, and the light-heavy neutrino mixing is also given at the leading order, taking into account the higher-order terms would sophisticate the exploitation of common SBS in the neutrino sector, although these higher-order corrections do not significantly affect the observables considered. Given these observations and in order to simplify the analysis, we shall assume that the heavy neutrinos do not carry the hypothetical $U(1)$ charges in the physical mass-eigenstate basis, and set $\mathcal{N}=0$ in the transformation rules defined by Eq.~\eqref{symmetry transformation}. In this case, the same criteria lead us to build the neutrino Yukawa matrices with only the rows specified, $(\tilde{Y}_{\nu})_{ij},(\mathcal{Y}_{\nu})_{ij} \propto y^\nu_i$, where $y^\nu_i$ denote the dimensionless Yukawa eigenvalues of the effective neutrino mass matrix. The $j$-columns of $(\tilde{Y}_{\nu})_{ij}$ and $(\mathcal{Y}_{\nu})_{ij}$ need not be specified, because the right-handed counterparts do not participate in the symmetry violation. Nevertheless, the effective active neutrino mass matrix in the non-physical basis can still exhibit a texture like  Eq.~\eqref{mp2} due to the seesaw formula specified by Eq.~\eqref{seesaw}, which is necessary to guarantee the synchronous vanishing of neutrino mixing via Eq.~\eqref{synchronization}. In this way, the hypothetically vanishing of charged lepton and active neutrino masses (or equivalently their respective Yukawa eigenvalues) renders a universal $U_L(1)^{3}$ symmetry restoration throughout the Yukawa Lagrangian (Eq.~\eqref{lag}), as well as in the $W^\pm$-mediated charged currents.

In the current work, as we are mainly concerned with the effects of additional Yukawa interactions on explaining the NP signals considered, we shall focus only on $\mathcal{Y}_f$ but without delving into the flavor structures of Eq.~\eqref{mp2}.  Eq.~\eqref{mp}, on the other hand, can stir up even richer phenomenologies, as such a structure indicates that, for the additional Yukawa interactions, the flavor-specific couplings $y^f_i$ can be either enhanced or suppressed by the non-universal powers $\tilde{n}_i$. Here we are interested in how large the tree-level FCNC couplings of the scalars to the lighter flavors can be allowed when those couplings to the heavier flavors are responsible for the NP signals considered. To this end, we propose a simple mass-powered parametrization based on Eq.~\eqref{mp} by scaling the SBS with a common power and a dimensional normalization factor,
\begin{equation}\label{mass-powered}
(\mathcal{Y}_{f})_{ij} =
\begin{cases} \left( \frac{m_i \times m_j}{\Lambda^{2}}\right)^n\,e^{-i\theta_{ij}}, \quad f=u, d, \ell,
 \\[0.2cm]
 \left(\frac{m_i\times M_i}{\Lambda_{j}^{2}}\right)^n\,e^{-i\theta_{ij}}, \quad f=\nu,
\end{cases}
\end{equation}
where $\Lambda$ denotes the dimension normalization factor, $n$ is the power, and $\theta$ represents the possible $CP$-violating phase. Note that, as $\mathcal{Y}_\nu$ is not specified in the $j$-columns, we have phenomenologically introduced the right-handed neutrino masses $M_i$, with the index collocation being different from the Dirac fermions. In addition, the dimension normalization factors in the neutrino sector $\Lambda_j$ can be traced back to  two origins of heavy neutrinos: TeV-scale heavy neutrinos and keV-scale sterile neutrino DM, as will be shown in the subsequent discussions. It should also be mentioned that the index collocation in $\mathcal{Y}_\nu$ is motivated by the explanation of $R_{K^{(*)}}$ anomalies under the $\ell_i \to \ell_j \gamma$ constraints, as will be clarified later. 
Finally,  Eq.~\eqref{mass-powered} indicates that the FCNC mediated by the neutral (pseudo)scalars ($H, A$) would be controlled by the fermion mass hierarchies. This can provide, therefore, a compelling suppression of FCNC in the lighter-flavor sector while allowing for sizable FCNC to explain the B-meson anomalies. 

\textit{Phenomenology}.---We now apply the mass-powered Yukawa textures to the neutrino mass, DM, as well as the $R_{K^{(*)}}$ and $R_{D^{(*)}}$ anomalies. In addition, we calculate the NP contribution to the muon $g-2$ as a further application. For numerical analyses, we set $\Lambda_u=m_t$, $\Lambda_\ell=m_\tau$ and $\Lambda_d\gg m_b$ to allow sizable top- and tau-associated Yukawa couplings, which are found to be essential for the LFUV explanations and, meanwhile, to suppress the FCNC in the down-quark sector. 
On the other hand, possible $CP$-violating phases in Eq.~\eqref{mass-powered} will be neglected unless stated otherwise. The scalar masses are considered around $M_{S}\simeq \mathcal{O}(500)~{\rm GeV}$ ($S=H^\pm, H, A$), to sufficiently alleviate the scalar effects on electroweak precision observables~\cite{Haller:2018nnx}.

\textit{Neutrino mass and DM}: As the neutrino mass spectrum is currently unknown, we cannot apply Eq.~\eqref{mass-powered} directly to the neutrino sector. It is known that, to have a stable neutrino mass generation via a low-scale seesaw mechanism, a lepton-like $U(1)$ symmetry is usually considered in the flavor basis. In Ref.~\cite{Li:2018rax}, the lepton-like charges in the (primed) flavor basis are chosen as
\begin{align}
L_{N^\prime_{R1}}=0,\quad L_{N^\prime_{R2}}=-L_{N^\prime_{R3}}=1,\quad
L_{E^\prime_{Li}}=L_{e^\prime_{Ri}}=1.
\end{align} 
There, the Yukawa couplings involving $N^\prime_1$ and $N^\prime_3$ break the lepton-like $U(1)$ symmetry, and $N^\prime_{2}$ and $N^\prime_{3}$ are nearly degenerate. After rotating to the physical mass-eigenstate basis in which the neutrino mixing information is encoded in the Dirac Yukawa matrix $\tilde{Y}_\nu$, as has been constructed in Eq.~\eqref{lag}, one can find that, if the lightest sterile neutrino (denoted as $N_1$) plays the role of keV DM, all the elements of the first column of $\tilde{Y}_\nu$ will be strongly suppressed by the cosmological X-ray observation (see Ref.~\cite{Boyarsky:2018tvu} for an updated review), via the $W^\pm$-mediated $N_1\to \nu \gamma$ decay~\cite{Bulbul:2014sua,Boyarsky:2014jta,Boyarsky:2014ska}
\begin{align}\label{neutrino radiative decay}
\Gamma(N_1\to \nu \gamma) =\frac{9\sqrt{2} \alpha_{\rm em} G_F}{1024 \pi^4} \sum_{i=1}^3 \vert (\tilde{Y}_\nu)_{i1} \vert^2 M_1^3,
\end{align}
where $G_F$ and $\alpha_{\rm em}$ are the Fermi and the fine-structure constants, respectively.
Note that, we have expressed the decay width in terms of the Dirac Yukawa matrix, so that the usual constraint on the light-heavy neutrino mixing angle $\theta_\nu$ can be translated onto $(\tilde{Y}_\nu)_{i1}$ by the relation
\begin{align}\label{angle-Yukawa}
\theta^2_\nu\simeq\frac{\sqrt{2}}{4G_FM_1^2} \sum_{i=1}^3 \vert (\tilde{Y}_\nu)_{i1} \vert^2.
\end{align}
 In this case, only two neutrino masses will be generated at the atmospheric scale $\Delta m_{{\rm{atm}}}\simeq 0.05~{\rm eV}$~\cite{Esteban:2016qun}, and the resulting mass hierarchy assumes either the normal ordering (NO) or the inverted ordering (IO), 
\begin{align}
\text{NO}: 0\simeq m_1\ll m_2<m_3, \quad
\text{IO}: 0\simeq m_3\ll m_1<m_2.
\end{align}

As the DM candidate, the lightest sterile neutrino $N_1$ would be a Feebly Interacting Massive Particle~\cite{Hall:2009bx}, since the associated Yukawa couplings that break the lepton-like $U(1)$ symmetry are expected to be suppressed. On the other hand, as the corresponding  Dirac Yukawa coupling $\tilde{Y}_\nu$ is severely constrained, the relic density of $N_1$ will then be accumulated by the freeze-in mechanism~\cite{Hall:2009bx} via the decays $S\to N_1 + l$~($l=\ell$ or $\nu$), when the scalars are in thermal equilibrium with the SM bath. The relic abundance is then given by~\cite{Hall:2009bx} 
\begin{align}
\Omega_{N_1}h^2&\simeq  \frac{1.09 \times 10^{27} M_1}{32\pi g^{*3/2}}\sum_{i=1}^{3}\vert (\mathcal{Y}_{\nu})_{i1}\vert^2
\nonumber \\
&\times \left(M_A^{-1}+M_H^{-1}+2M_{H^\pm}^{-1}\right) ,
\end{align}
where $g^*=110.75$ denotes the  effective number of  relativistic degrees of freedom in the thermal bath around freeze-in temperature $T\sim M_S$, after taking into account the additional Higgs bosons besides those in the SM (the heavy Majorana neutrinos are not included and the active neutrinos are of Majorana type).

If $N_1$ is the source of  $3.5~{\rm keV}$ X-ray line~\cite{Kusenko:2009up},  then $M_1=7.1~{\rm keV}$  is fixed. In this case, the  active neutrino mass spectrum is  determined by
 \begin{align}\label{neutrino mass}
\text{NO}&: m_1\simeq 10^{-6}~{\rm eV},~
m_2=0.009~{\rm eV},
m_3=0.05~{\rm eV},
\nonumber \\[0.2cm]
\text{IO}&: m_3\simeq 10^{-6}~{\rm eV}, ~m_1=0.049~{\rm eV}, ~m_2=0.05~{\rm eV}.
\end{align}
On the other hand, the $7.1$~keV $N_1$ populates the warm DM region~\cite{Merle:2013wta,Adulpravitchai:2014xna}, and the explanation of the $3.5~{\rm keV}$ X-ray line via the radiative decay $N_1\to \nu \gamma$ can also be consistent with the Lyman-$\alpha$ observation~\cite{Merle:2014xpa}.

\begin{figure}[ht]
	\centering	
	\includegraphics[width=0.40\textwidth]{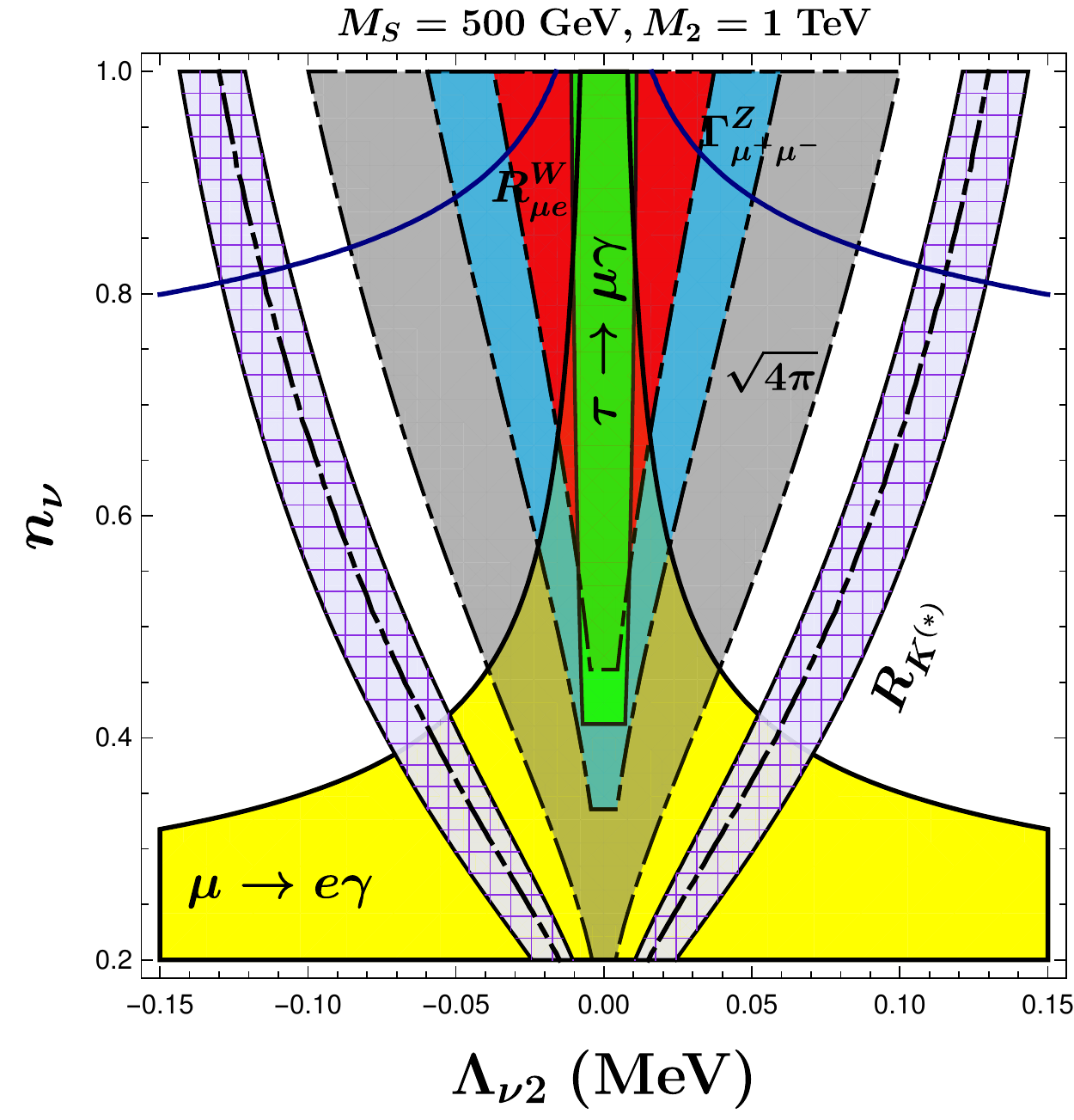}
	\caption{Explanation of $R_{K^{(*)}}$ anomalies in light of the updated LFUV global fit (shaded band)~\cite{Aebischer:2019mlg,Ciuchini:2019usw,Alguero:2019ptt,Arbey:2019duh,Alok:2019ufo}, with the black line corresponding to the best-fit point. The yellow and green regions are excluded by the $\mu \to e\gamma$ and $\tau \to \mu\gamma$ constraints, respectively. The light blue region is excluded by $Z\to \mu^+\mu^-$ and the red one by $R^W_{\mu e}$. Finally, the gray region is excluded by the perturbative unitarity requirement $\mathcal{Y}>\sqrt{4\pi}$. The blue line corresponds to the DM relic density $\Omega_{DM} h^2=0.12$~\cite{Tanabashi:2018oca} with a numerical hierarchy $\Lambda_{\nu 1}=10^5\Lambda_{\nu 2(3)}~{\rm MeV}$. }
	\label{RKfits}
\end{figure}

In Fig.~\ref{RKfits}, the current DM relic density, $\Omega_{N_1} h^2=0.12$~\cite{Tanabashi:2018oca}, is fitted with a numerical hierarchy $\Lambda_{\nu 1}=10^5 \Lambda_{\nu 2(3)}~{\rm MeV}$ (blue line). Here the choice of the hierarchy between $\Lambda_{\nu 1}$ and $\Lambda_{\nu 2(3)}$, as well as the approximation $\Lambda_{\nu 2}\simeq \Lambda_{\nu 3}$ is sensible, as $\Lambda_{\nu 1}$ corresponds to the DM interactions and the small difference $\Lambda_{\nu 2}- \Lambda_{\nu 3}$ can be treated as the breaking source of the lepton-like symmetry.  In addition, we have taken $M_2>M_{S}$ so as to open the decay channel $N_{2,3}\to S +l$, which can sufficiently decrease the lifetime of $N_{2,3}$. The resulting impact of $N_{2,3}$ on the effective relativistic degrees of freedom at the Big Bang Nucleosynthesis epoch can be, therefore, neglected safely~\cite{Hernandez:2014fha}.

\textit{$R_{K^{(*)}}$}:  The dominant NP contribution to $b\to s\ell^+\ell^-$ stems from the box diagrams mediated by charged Higgs boson, top quark and heavy neutrino, which result in the effective Wilson coefficients:
\begin{align}
C_{9\ell}^{\rm NP}&=-C_{10 \ell}^{\rm NP}
\nonumber \\
 &=\frac{G_F^{-1}\alpha_{\rm em}^{-1}}{64\sqrt{2}\pi M_W^{2}} \sum_{i=1}^3\vert (\mathcal{Y}_{\nu})_{\ell i}\vert^2 \; \vert (\mathcal{Y}_{u})_{33}\vert^2 \;\mathcal{G}(\lambda_{\pm},\lambda_i,\lambda_t),
\end{align} 
corresponding to the semi-leptonic operators
\begin{align}
\mathcal{O}_{9(10)}=\frac{\alpha_{\rm em}}{4\pi} \left(\bar{s} \gamma_{\mu}P_L b\right)\left(\bar{\ell} \gamma^{\mu}(\gamma_5) \ell \right). 
\end{align} 
Here, $\lambda_{\pm}=M_{H^\pm}^2/M_W^2$, $\lambda_{i}=M_{i}^2/M_W^2$, $\lambda_{t}=m_t^2/M_W^2$, and the loop function
\begin{align}
\mathcal{G}(x,y,z)&=\frac{y^2 \log \left(x/y\right)}{(x-y)^2 (y-z)}+\frac{z^2 \log
	\left(x/z\right)}{(x-z)^2 (z-y)}
\nonumber \\
&-\frac{x}{(x-y) (x-z)}.
\end{align} 
It is seen that, due to the neutrino mass hierarchies encoded in the mass-powered parametrization~(see Eq.~\eqref{mass-powered}), the necessary flavor non-universal couplings appear naturally, and the LFUV observables $R_{K^{(*)}}$ can be explained by the fact that $M_{2(3)}\gg M_1$ and $m_2>m_1$. Nevertheless, such a muon-neutrino specific Yukawa coupling may give significant effects on muon-associated observables. To clarify that a successful explanation of the $R_{K^{(*)}}$ anomalies is possible with such a coupling, we further take into account the primary constraints from $\mu \to e\gamma$, $\tau\to\mu\gamma$, $Z\to \mu^+\mu^-$, as well as the ratio $R^W_{\mu e}\equiv \Gamma(W\to \mu \nu)/\Gamma(W\to e \nu)$ that tests the lepton-flavor universality~\cite{Tanabashi:2018oca}. Finally, the perturbative unitarity requirement $\mathcal{Y}>\sqrt{4\pi}$ is also imposed.

From Fig.~\ref{RKfits}, it is clearly seen that the updated model-independent analyses~\cite{Aebischer:2019mlg,Ciuchini:2019usw,Alguero:2019ptt,Arbey:2019duh} that prefer $C_{9\mu}^{\rm NP}=-C_{10 \mu}^{\rm NP}<0$ with a significance at the level of $\sim 4\sigma$ can be well reproduced here, as shown by the shaded band. Here we have taken the IO pattern of the active neutrino masses (see Eq.~\eqref{neutrino mass}). Note that, for the sake of plotting, we have compiled the possibly negative values of $\mathcal{Y}_\nu$ to that of $\Lambda_{\nu 2}$, since the considered observables depend on $\vert\mathcal{Y}_\nu\vert^2$. With such a numerical setup, the resolution of the $R_{K^{(*)}}$ anomalies exists under all the constraints considered. However, we find that if the active neutrino masses have the NO pattern, the parameter region allowed by the $R_{K^{(*)}}$ explanation will be completely ruled out by the $\tau \to \mu \gamma$ constraint.

Furthermore, we have also numerically confirmed that, if $\mathcal{Y}_\nu$ exhibits a texture similar to that of the Dirac fermions, 
\begin{align}
(\mathcal{Y}_{\nu})_{ij}\propto \frac{m_i^n \times M_j^n}{\Lambda_j^{2n}},
\end{align} 
rather than the one proposed in Eq.~\eqref{mass-powered}, the resolution of the $R_{K^{(*)}}$ anomalies would be excluded either by $\mu \to e\gamma$ or by $\tau \to \mu\gamma$, no matter which hierarchies of the active neutrino masses are taken. Therefore, being in association with the lightest heavy neutrino as a $7.1~{\rm keV}$ DM, the mass-powered texture of $\mathcal{Y}_\nu$ in Eq.~\eqref{mass-powered} serves a twofold role, on the one hand accounting for the LFUV in $R_{K^{(*)}}$ and on the other hand predicting an IO pattern of the active neutrino masses.

\textit{$R_{D^{(*)}}$}:  The NP effect on $b\to c\ell_i \nu$ transitions arises  from the tree-level charged-Higgs contribution. Neglecting the suppressed down-quark Yukawa couplings, the effective Hamiltonian  is given by 
\begin{align}
\mathcal{H}_{\rm eff}=\frac{(\mathcal{Y}_{\ell})_{ii}(\mathcal{Y}^*_u)_{32}}{M_{H^\pm}^{2}}\,V_{tb}\,(\bar{c}P_Lb)(\bar\ell_i P_L \nu).
\end{align}
It can be readily seen that, the lepton-flavor non-universality is induced by  mass hierarchy encoded in  $\mathcal{Y}_\ell$. For numerical analysis,
 we implement the Wilson coefficient to  the updated formulae in Ref.~\cite{Blanke:2018yud}. Besides, to verify that our model can provide a feasible resolution of the $R_{D^{(*)}}$ anomalies, we consider the bounds from the branching ratio $\mathcal{B}(B_c\to \tau \nu)$ derived from the $B_c$ lifetime~\cite{Alonso:2016oyd,Celis:2016azn,Li:2016vvp}, and the $D^*$ longitudinal polarization fraction~\cite{Abdesselam:2019wbt}, 
\begin{align}
F_L^{D^*}=0.60\pm 0.08(\text{stat})\pm 0.04(\text{syst}),
\end{align}
which differs from its SM prediction~\cite{Hu:2018veh},
\begin{align}
F_L^{D^*}=0.455\pm 0.003,
\end{align}
 by $\sim1.6\sigma$. Following Ref.~\cite{Li:2018aov}, we also consider the mass difference $\Delta M_s$ in the $B_s-\bar{B}_s$ system and the inclusive $B\to X_s \gamma$ branching ratio, both of which receive sizable NP effects from the top-associated Yukawa couplings.

\begin{figure}[ht]
	\centering	
	\includegraphics[width=0.40\textwidth]{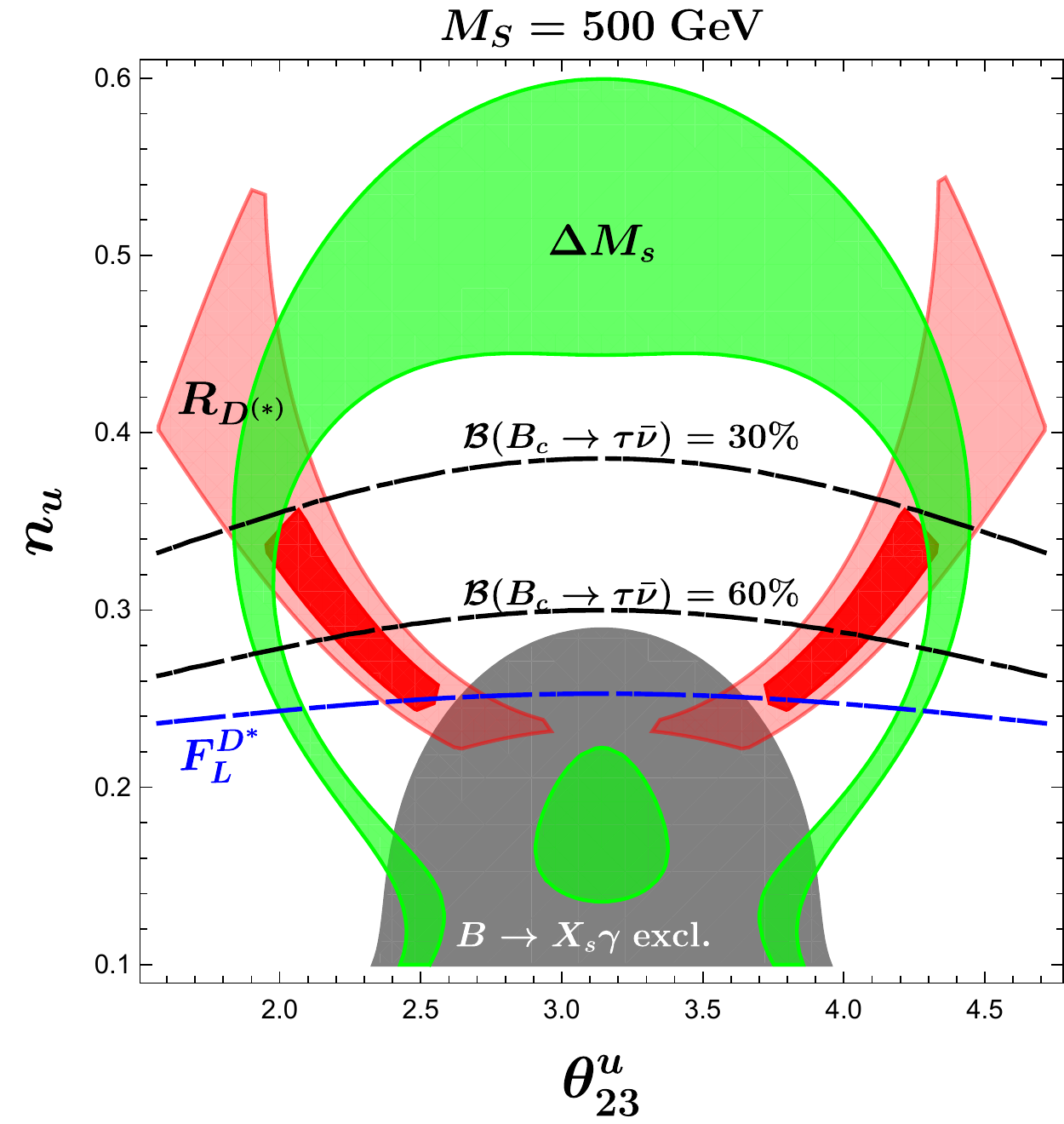}
	\caption{A simultaneous explanation of $R_D$ and $R_{D^*}$ anomalies is shown in red with $1\sigma$ (darker) and $2\sigma$ (lighter) experimental errors, respectively. The regions below the black and blue curves are already excluded by $\mathcal{B}(B_c\to \tau \nu )>30\%$, $\mathcal{B}(B_c\to \tau \nu )>60\%$, and $F_{L}^{D^*}$ (at $1\sigma$ level). The gray region is excluded by $\mathcal{B}(B\to X_s \gamma)$, while the  $2\sigma$ range allowed by $\Delta M_s^{\rm exp}/ \Delta M_s^{\rm SM}$ is indicated by the green region.}
	\label{RDfits}
\end{figure}

As shown in Fig.~\ref{RDfits}, a $1\sigma$-level explanation of $R_{D^{(*)}}$ anomalies is feasible under the constraint $\mathcal{B}(B_c\to \tau \nu)<60\%$~\cite{Blanke:2018yud,Blanke:2019qrx}, while a simultaneous explanation can be realized only at $2\sigma$ level if $ \mathcal{B}(B_c\to \tau \nu)<30\%$~\cite{Alonso:2016oyd} is imposed. Here the $2\sigma$ range allowed by $\Delta M_s^{\rm exp}/\Delta M_s^{\rm SM}$ (green region) is obtained with $\Delta M_s^{\rm exp}=(17.757\pm 0.021)~\rm ps^{-1}$~\cite{Amhis:2019ckw} and $\Delta M_s^{\rm SM}=(20.01\pm 1.25)~\rm ps^{-1}$~\cite{DiLuzio:2017fdq,Aoki:2019cca}. On the other hand, by taking $n_\ell\simeq1$, the charged-lepton mass hierarchies encoded in $\mathcal{Y}_\ell$ would induce significant effects on the $\tau$ channel but suppress those in the $\mu/e$ modes. This ensures negligible effects on $\mathcal{B}(B\rightarrow D^{(\ast)} \mu \bar{\nu})/\mathcal{B}(B\rightarrow D^{(\ast)} e \bar{\nu})$~\cite{Jung:2018lfu} and, at the same time, provides a natural explanation of the $R_{D^{(*)}}$ anomalies.

Concerning the constraint from $H^\pm \to \tau \nu$ searches at the LHC, it is found that the decay width of $H^\pm$ is now dominated by $\mathcal{B}(H^\pm\to \tau \nu)$~($30\%$) and $\mathcal{B}(H^\pm\to tb)$~($70\%$), and the strength of cross section times branching ratio, $\sigma(pp\to H^\pm)\times\mathcal{B}(H^\pm \to \tau \nu)$, will be reduced compared to the case with $\mathcal{B}(H^\pm\to \tau \nu)\simeq 90\%$. Therefore, the compatibility between the $R_{D^{(*)}}$ explanation and the LHC constraint from $H^\pm \to \tau \nu$ searches~\cite{Iguro:2018fni} can be realized, as the latter becomes weaker by at least a factor of three.

\textit{$\Delta a_\mu$}: As a further application of the mass-powered parametrization, let us now consider the longstanding puzzle observed in the muon $g-2$, $a_\mu=(g_\mu-2)/2$. The current value~\cite{Keshavarzi:2018mgv},
\begin{align}
\Delta a_\mu =a^{\rm exp}_\mu-a^{\rm SM}_\mu=(27.06\pm7.26) \times 10^{-10},
\end{align} 
exhibits a $3.7\sigma$ discrepancy between theory and experiment. However, the present SM prediction is still plagued by large hadronic uncertainties, leading to a possible range of $0.7-4.2\sigma$ deviations~\cite{Jegerlehner:2009ry}. Furthermore, a milder discrepancy, including the no-NP solution, has also been implied by the recently refined lattice calculations of the hadronic contributions to the muon $g-2$~\cite{Borsanyi:2017zdw,Blum:2018mom,Davier:2019can}.

\begin{figure}[ht]
	\centering	
	\includegraphics[width=0.40\textwidth]{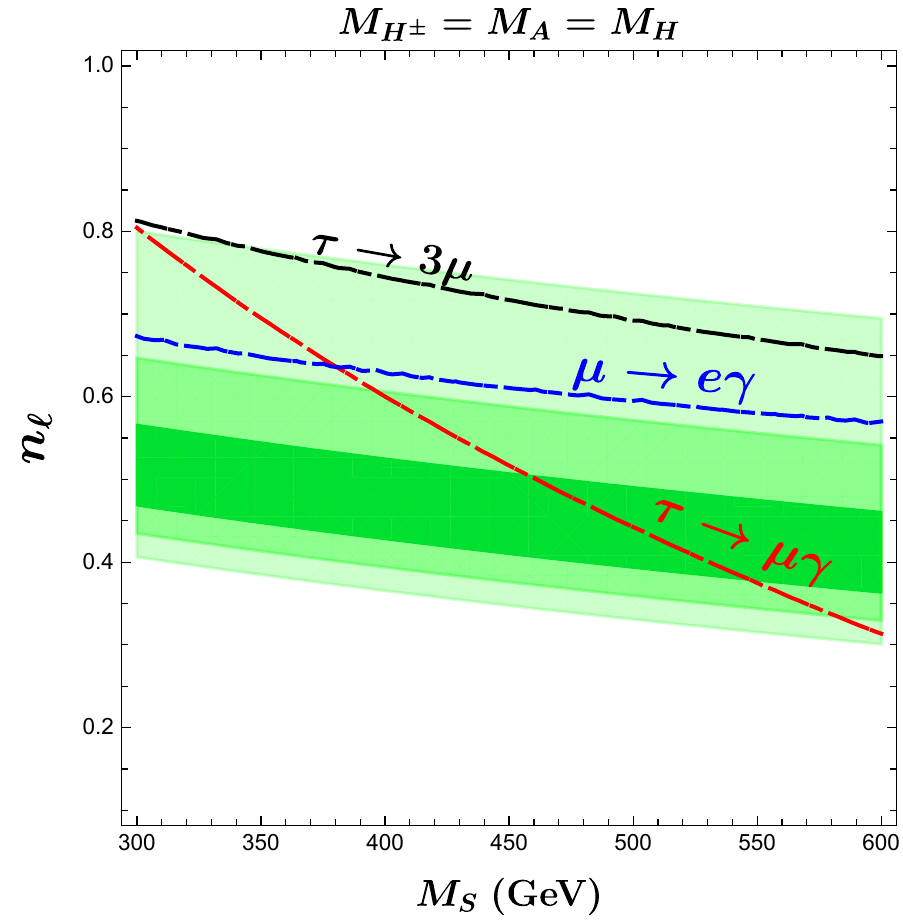}
	\caption{NP effects on $\Delta a_\mu$ under the constraints from LFV processes. The $1-3\sigma$ regions of $\Delta a_\mu$~\cite{Keshavarzi:2018mgv} are shown in green. The region below the black, blue, and red curves are excluded by $\tau \to 3\mu$, $\mu \to e \gamma$ and $\tau \to  \mu \gamma$, respectively.  $\theta^u_{33}=\pi$  is chosen to obtain a maximal NP contribution to $\Delta a_\mu$.}
	\label{amufits}
\end{figure}
Here, the NP contributions arise from both the one-loop diagrams involving the muon-tau Yukawa couplings,  as well as the two-loop Barr-Zee diagrams involving the top and tau Yukawa couplings even in the degenerate mass regime of the scalar bosons~\cite{Li:2018aov}. To sum over these contributions, we have taken the formulae~\cite{Li:2018aov} by rescaling the Yukawa couplings with the mass-powered parametrization given in Eq.~\eqref{mass-powered}. In this regime, we show, in Fig.~\ref{amufits}, the parameter regions required to explain the current $g-2$ data, as well as the constraints from lepton-flavor changing (LFV) processes, with the primary one resulting from $\tau \to 3\mu$. It is seen that the mass-powered parametrization cannot explain the current $3.7\sigma$ discrepancy of the muon $g-2$ at $1\sigma$ level. Instead, the $3\sigma$-level accommodation suggests a milder discrepancy or even no-NP solution of the muon $g-2$, which has also been hinted by the latest refined lattice results~\cite{Borsanyi:2017zdw,Blum:2018mom,Davier:2019can}.

\textit{Conclusions}.---We have considered an interesting scenario where the Yukawa interactions are constructed by a \textit{symmetry violation} guideline, rather than by the usually adopted \textit{symmetry invariance} principle. Implementing such a guideline into the 2HDM$+3N_R$ framework, we found that the model becomes very constrained but its predictive power is enhanced dramatically. By attributing the SBS of a hypothetical $U_Q(1)^3\times U_L(1)^3$ symmetry to the known fermion masses (or equivalently the dimensionless Yukawa eigenvalues), we found that the unknown Yukawa matrices of the model can be built out of the known fermion masses. As a phenomenological application, we have investigated the interplay between the severely constrained light-flavor FCNCs and the explanations of NP signals from heavy-flavor FCNCs by proposing a simple mass-powered parametrization of the additional Yukawa matrices.

The mass-powered texture renders a correlative incorporation of the neutrino mass, DM, as well as the LFUV observed in $R_{K^{(*)}}$ and $R_{D^{(*)}}$. Let us summarize the main results. Two atmospheric-scale neutrinos are generated by a $U(1)$-protected low-scale seesaw mechanism, while the remaining much lighter one is fixed by a $7.1~{\rm keV}$ sterile neutrino DM. Such a DM can explain the $3.5~{\rm keV}$ X-ray line, and its relic density is produced by the freeze-in mechanism via the thermalized scalar decays. In light of the mass-powered parametrization, the IO pattern of the active neutrino masses, $0\simeq m_3\ll m_1<m_2$, together with a strong hierarchical heavy neutrino masses, $M_1\ll M_2\simeq M_3$, accounts for the $R_{K^{(*)}}$ anomalies, while the charged-lepton mass hierarchies explain the $R_{D^{(*)}}$ data. Finally, a milder discrepancy of the muon $g-2$ suggested here can be served as a future probe for the particular parametrization.

\vspace{0.1cm}
This work is supported by the National Natural Science Foundation of China under Grant Nos.~11675061 and 11435003, as well as by the Fundamental Research Funds for the Central Universities under Grant Nos.~CCNU18TS029 and 2019YBZZ079.

\bibliographystyle{apsrev4-1}
\bibliography{reference}

\end{document}